# Polarized phonons carry the missing angular momentum in femtosecond demagnetization

S. R. Tauchert[1,2], M. Volkov[1,2], D. Ehberger[2], D. Kazenwadel[1], M. Evers[1], H. Lange[1], A. Donges[1], A. Book[3], W. Kreuzpaintner[3,4,5], U. Nowak[1], P. Baum[1,2]

[1] *Universität Konstanz, Fachbereich Physik, 78464 Konstanz, Germany*

[2] *Ludwig-Maximilians-Universität München, Am Coulombwall 1, 85748 Garching, Germany*

[3] *Technische Universität München, Physik-Department E21, James-Franck-Str. 1, 85748 Garching, Germany*

[4] *Institute of High Energy Physics, Chinese Academy of Sciences (CAS), Beijing 100049, China*

[5] *Spallation Neutron Source Science Center, No. 1 Zhongziyuan Road, Dalang, Dongguan, 523803, China*

\**peter.baum@uni-konstanz.de*

**Magnetic phenomena are ubiquitous in our surroundings and indispensable for modern science and technology, but it is notoriously difficult to change the magnetic order of a material in a rapid way. However, if a thin nickel film is subjected to ultrashort laser pulses, it can lose its magnetic order almost completely within merely femtosecond times [1]. This phenomenon, in the meantime also observed in many other materials [2-7], has connected magnetism with femtosecond optics in an efficient, ultrafast and complex way, offering opportunities for rapid information processing [8-12] or ultrafast spintronics at frequencies approaching those of light [8,9,13]. Consequently, the physics of ultrafast demagnetization is central to modern material research [1-7,14-28], but a crucial question has remained elusive: If a material loses its magnetization within only femtoseconds, where is the missing angular momentum in such short time? Here we use ultrafast electron diffraction to reveal in nickel an almost instantaneous, long-lasting, non-equilibrium population of anisotropic high-frequency phonons that appear as quickly as the magnetic order is lost. The anisotropy plane is perpendicular to the direction of the initial magnetization and the atomic oscillation**

**amplitude is 2 pm. We explain these observations by means of circularly polarized phonons that quickly absorb the missing angular momentum of the spin system before the slower onset of a macroscopic sample rotation. The time that is needed for demagnetization is related to the time it takes to accelerate the atoms. These results provide an atomistic picture of ultrafast demagnetization under adherence to all conservation laws but also demonstrate the general importance of polarized phonons for non-equilibrium dynamics and provide innovative ways for controlling materials on atomic dimensions.**

Since its discovery in 1996 [1], the physics of ultrafast demagnetization has been examined with a wide variety of measurement techniques [1-7,14-28] for clarifying the interplay of laser photons, electrons, spins and lattice dynamics for the reaction path. Pioneering works have suggested spin transport [7,17,18], ultrafast strain waves [22] or an efficient spin lattice coupling [2,15,20,25,26] as potential mechanisms for the flow of angular momentum. However, vivid arguments are also available against each of those pictures. For example, Eschenlohr et al. [18] have demonstrated the significance of spin transport in metal heterostructures while Dornes et al. [22] have reported x-ray diffraction data on iron that indicate the role of mechanical torques and strain waves via an ultrafast Einstein-de Haas effect. Atoms move collectively in this picture, in contrast to earlier indications of a transient lattice disorder [15]. In strong laser fields, spin polarization can be steered away quickly [13], but at normal excitation strengths spin transport is not substantial [23]. Stamm et al. [21] have provided pioneering evidence with x-ray magnetic circular dichroism on the probable role of lattice dynamics and Dürr et al. [14] have discovered a stepwise phonon dynamics, but Chen et al. [24] have denied the relevance of phonon dynamics expect for their later participation in the back reaction. A comprehensive picture of the reaction path under conservation of angular momentum is therefore not available, although it touches the foundations of magnetism and spintronics and may help to improve the efficiency, speed or versatility of devices for rapid data storage and manipulation.

**Experiment**

Here we report the use of ultrafast electron diffraction [29,30] with terahertz-compressed electron pulses [31] as a direct probe of lattice dynamics in space and time. Space charge effects in the ultrashort electron pulses are avoided by using less than 20 electrons per pulse. Figure 1a depicts the experiment. A single-crystalline layer of nickel with a thickness of 22 nm is epitaxially grown on a single-crystalline silicon membrane with a hydrogen-terminated (100)-surface [32]. A 6-nm thick copper seed layer improves the crystal growth (see methods). Figure 1b depicts the orientation of the epitaxial growth; Ni-(100) attaches to Si-(100) at an angle of 45°. Demagnetization is triggered by femtosecond laser pulses with a center wavelength of 1030 nm and a pulse duration of ~90 fs (see methods). Although the nominal absorption depth of nickel is ~15 nm, multi-layer interferences produce an almost homogeneous excitation density along the $z$ axis (see methods). The excitation energy density (fluence) at the beam center is set to ~4 mJ/cm$^2$ of which ~1.5 mJ/cm$^2$ are absorbed, conditions for which nickel demagnetizes by ~50 % within ~250 fs [15,28]. Magneto-optical Faraday rotation measurements under similar excitation conditions as in the diffraction experiment indeed show an ultrafast and long-lived demagnetization of 40-50% (extended data Fig. S13). Femtosecond electron pulses are generated by two-photon photoemission from a gold cathode and accelerated to a kinetic energy of 70 keV, providing a de Broglie wavelength of ~4 pm. A laser-generated single-cycle terahertz field is used for a temporal compression of the electron pulses down to a duration of ~90 fs [31], measured by all-optical streaking (see Fig. 1c). The pump-probe repetition rate is 25-50 kHz and the electron beam on the specimen is about two times smaller than the excitation beam (see methods). After each femtosecond laser pulse, the sample cools back to ambient temperature within ~50 K, evident from measurements of almost equal Bragg spot intensities before time zero as compared to an experiment with no laser excitation at all (see methods). We infer a base temperature of the experiment of 340±50 K.

In order to predefine an initial in-plane magnetization direction before each pump-probe cycle, we use a series of permanent ring magnets as depicted in Fig. 1a. The first magnet provides at a distance of 2±1 mm from its front surface an approximately homogeneous magnetic field of 22±8 mT with a direction in plane of the specimen (*xy*-plane; see Fig. 1a). The second magnet serves for creating an opposing magnetic field that cancels the off-axis deflection of the electron beam by the Lorentz forces of the first magnet. In this way, the electron beam direction and diffraction pattern remain almost undistorted. The whole magnetic assembly can be rotated around the *z*-axis in order to study time-dependent Bragg diffraction as a function of the absolute magnetic field direction along the *x*-axis or *y*-axis, that is, the [100] or [010] direction of the nickel crystal (see Fig. 1b). The diffraction pattern is imaged onto a single-electron sensitive detector system (F416, TVIPS GmbH) with a magnetic solenoid lens with calibrated image rotation. Consequently, the measured Bragg spots on the screen are linked to the crystallographic orientation of the sample and the absolute direction of the initial magnetization.

**Results**

Figure 1d depicts the static electron diffraction pattern with no laser excitation. We see a series of nickel Bragg spots at the expected diffraction angles (dotted circle). Some silicon spots from the support membrane, for example (220) and ($\bar{2}\bar{2}0$), are visible at slightly smaller diffraction angles (see Fig. 1f), because the unit cell of nickel in [100] direction is ~8 % smaller than the unit cell of silicon in [110] direction. The nickel crystal therefore adapts to the lattice of the silicon substrate by dislocations. Figure 1e depicts a rocking curve, that is, the integrated Bragg spot intensity as a function of specimen tilt. We see a width of ~6.5°, indicating about ±2 atomic displacements over the layer thickness along the $z$ direction. The faint ring inside of the dashed line and the four very weak spots at 45° are caused by nickel's NiO$_x$ termination (see methods).

After femtosecond laser excitation, the measured diffraction pattern changes due to structural dynamics of the lattice. We see within the measured delay range of several picoseconds no measurable broadening of the Bragg spots and no displacements (extended data Fig. S13), indicating the absence of transient strain or lattice expansion on such time scale. However, there are substantial changes of the Bragg spot intensities of nickel, linked to structural dynamics within the unit cell. Figure 1g shows a close-up of diffraction difference data at the example of the Ni-(020) spot (dashed region) and its surroundings. Time-dependent intensity changes $I_{hkl}(t)$ are obtained by reference to the measured intensity before time zero, that is, $I_{hkl}(t) = I_{hkl}^{raw}(t)/I_{hkl}^{raw}(t<0)$ where $I_{hkl}^{raw}$ denotes the integrated digital counts in the area of the spot (see Fig. 1f). In this way, any systematic sensitivity variations of the apparatus cancel out. Additionally, we normalize $I_{hkl}(t)$ to $I_{000}(t)$ in order to compensate for potential temporal drifts of the electron beam. Figure 2a shows the resulting data for the symmetry-equivalent Bragg spots $I_{\{200\}} = \frac{1}{4}(I_{200} + I_{020} + I_{\bar{2}00} + I_{0\bar{2}0})$ and $I_{\{220\}} = \frac{1}{4}(I_{220} + I_{\bar{2}20} + I_{2\bar{2}0} + I_{\bar{2}\bar{2}0})$ as a function of time. We see that the elevated lattice temperature after laser excitation diminishes the Bragg spots via the Debye-Waller effect. The effective electron-phonon coupling time [33] is ~700 fs (solid line) and the measured ratio $(1 - I_{\{220\}})/(1 - I_{\{200\}}) \approx 2$ matches the expected picture of a random disorder in the lattice in form of a transient temperature increase [1,15,33].

However, additional anisotropic phonon distributions are revealed by an asymmetry analysis of pairs of Bragg spots that have crystallographic equivalence but differ in their orientation with respect to the magnetic field. In particular, we compare (200) and ($\bar{2}$00) with (020) and (0$\bar{2}$0), that is, the upper and lower vs the left and right Bragg spots on the circle in Fig. 1d. Any lattice disorder in form of a transient temperature can only affect all these four spots in the same way, and differences in their dynamics therefore reveal a nonthermal phonon dynamics in violation of equipartition. We note that the selected way of averaging opposing Bragg spots makes the experiment particularly insensitive to potential changes of the crystallographic axes. Spots (200) and ($\bar{2}$00) as well as (020) and (0$\bar{2}$0) are Friedel pairs and for a symmetric diffraction pattern their sum is insensitive to tilt or shear of the lattice (see methods). Also, the rocking curve shows some residual broadening (see Fig. 1e). We therefore probe exclusively sub-unit-cell dynamics and atomistic disorder, not mechanical strain waves or torques.

Figure 2b shows the time-dependent averaged intensity of (200) and ($\bar{2}$00) divided by the time-dependent average intensity of (020) and (0$\bar{2}$0). We see distinct asymmetric changes after time zero depending on the orientation of the initial magnetic field, symbolized by the green arrows. Spots (200) and ($\bar{2}$00) become more intense than (020) and (0$\bar{2}$0) if the initial magnetic field is set along nickel's [100] direction but the ratio reverses for an initial magnetization along the [010] axis. The time it takes to develop this anisotropy is <500 fs (dotted lines), likely ~200 fs (solid lines), and there is no detectable onset of a decay within the measured delay range. The anisotropic effect amounts to roughly 25% of the isotropic phonon disorder.

These results show that laser excitation of pre-magnetized nickel not only causes an ultrafast increase of lattice temperature in form of isotropic atomic displacements (see Fig. 2a), but also an additional, long-lived and anisotropic lattice motion relative to the direction of the initial magnetization (see Fig. 2b). This result implies an ultrafast, non-thermal phonon dynamics with a direct magnetic origin beyond the isotropic dynamics that has been observed before [14]. In other words, spin couples quickly and directly to symmetry-breaking high-frequency phonons with an oscillation in real space in perpendicular direction to the initial magnetization. This result is a direct observation of ultrafast spin-phonon coupling that has so far only been supposed [1] or indirectly been inferred [21].

**Interpretation**

We argue in the following that the origin of our observations is rotational atomic motions with angular momentum, that is, phonons with a circular polarization [34-36]. Figure 2c depicts such dynamics: atoms move around their equilibrium position with a rotation axis that is aligned with the magnetic field and with an average radius that is determined by the spin angular momentum that is lost by the demagnetization. Consequently, we perceive with our electron diffraction an anisotropic atomic disorder and a Debye-Waller effect only along such crystallographic directions that align perpendicularly to the initial direction of magnetization.

We support this interpretation by a nearly parameter-free consideration of the conservation laws in combination with molecular dynamics simulations. Prior to demagnetization, each nickel atom carries a magnetic moment of $\mu_{Ni} \approx 0.6\ \mu_B$, where $\mu_B$ is the Bohr magneton. Demagnetization by D ≈ 50 % (extended data Fig. S14) implies that each nickel atom must get rid of a spin angular momentum of $\Delta L = D \frac{\mu_{Ni} \hbar}{g_e \mu_B} \approx 0.15\ \hbar$, where $g_e \approx 2$ is the gyromagnetic ratio of the electron. No angular momentum can travel out of the excited region on femtosecond and picosecond time scales. If the lattice accepts this missing $\Delta L$ in form of rotational dynamics, each atom must move around its equilibrium position within the unit cell with an average radius of $R \approx \sqrt{\frac{\Delta L}{M_{Ni} \omega}}$, where $M_{Ni}$ is the mass of a nickel atom and $\omega$ is the angular frequency of the rotation, in order to conserve total angular momentum. The phonon spectrum of nickel has a peak in the density of states around $\omega/2\pi \approx 8$ THz [37] and with this assumption we obtain $R \approx 1.8$ pm. In electron diffraction, the intensity $I$ of a Bragg spot is related to random atomic displacements via the Debye-Waller factor

$$\frac{I}{I_0} = \langle \exp\left[\frac{2\pi i}{a}\begin{pmatrix}h\\k\\l\end{pmatrix}\cdot\begin{pmatrix}u_x\\u_y\\u_z\end{pmatrix}\right]\rangle^2 \qquad (1)$$

where $h, k, l$ are the Miller indices of the Bragg spot, $I_0$ is the intensity without disorder, $a \approx 352$ pm is the lattice constant of nickel, $u_{x,y,z}$ are the atomic displacements and $\langle\ \rangle$ denotes an averaging over the probed crystal volume. The inner product in Eq. (1) makes such Bragg spots with a zero in one of the Miller indices insensitive to atomic disorder along the corresponding dimension in real space. Hence, electrons 'see' only disorder in a direction that they diffract to. For example, the (200) and ($\bar{2}$00) spots are only sensitive to disorder along $x$, not $y$ or $z$. If the initial magnetization aligns with $y$, atoms are only displaced along $x$ and $z$ (see left panel in Fig. 2c).

Under the assumption of a circular trajectory with $u_x = R\cos(\phi)$, $u_z = R\sin(\phi)$ and $u_y = 0$, where $\phi$ is a supposedly random rotational phase for each atom in the laser-excited crystal volume, we obtain $I/I_0 \approx 0.998$ for Bragg spots (200) and ($\bar{2}$00) but $I/I_0 = 1$ for (020) and (0$\bar{2}$0). The experimentally measured ratio $I_{\{200\}}/I_{\{020\}}$ becomes 0.998. In contrast, for an initial magnetization along x (see right panel in Fig. 2c), we obtain $I_{\{200\}}/I_{\{020\}} \approx 1.002$. Despite its simplicity, this estimation reproduces the measurement results (see Fig. 2b) in magnitude and sign. We conclude that almost all of the angular momentum that is missing from ultrafast demagnetization arrives within <500 fs in the lattice in form of anisotropic circular motions of the atoms. Potential alternative explanations for our data are considered and invalidated in the methods section.

**Molecular dynamics simulations**

Molecular dynamics simulations support this interpretation. Using the simulation code LAMMPS [38], we set up a cubic nickel crystal in free space and use atomic pair interactions as calculated by the embedded-atom method [39]. First, we let the system relax and then, at time zero, we instantaneously create at randomly selected nickel atoms an angular momentum of $\hbar$ via a combination of local displacement and linear momentum; see methods. In this way, we excite the entire phonon spectrum at once and let the system decide which ones will prevail. Quantitatively, one quantum of $\hbar$ at every seventh nickel atom corresponds to the required amount of total angular momentum for a macroscopic demagnetization of 42 %. As a function of time, we then compute electron diffraction patterns [40] by averaging over up to 100 simulation runs. Different initial temperatures, system sizes and boundary conditions are inspected to infer the necessary convergences (see methods).

Figure 3 depicts the results. After laser absorption (see Fig. 3a), the imprint of angular momentum for individual atoms quickly develops into a sea of anisotropic high-frequency oscillations with a mean velocity distribution in violation of equipartition. Figure 3b depicts a histogram for all atoms in the simulation volume for the time range of the first 500 fs. The distribution is about two times wider along the x axis than along the z axis. Figure 3c shows the simulated Bragg spot anisotropy in analogy to the measured data of Fig. 2b. The different line styles denote an increasing number of nickel atoms in the simulated system, ~$10^4$ for the dotted line, ~$10^6$ for the dashed line and quasi-infinite with help of period boundary conditions for the solid line. Simulations with periodic boundary conditions avoid surface effects, although technically the total angular momentum is not conserved. All traces resemble the key feature of the experiment, namely the almost instantaneous onset of a long-lasting Bragg spot anisotropy. The coherent oscillations around time zero are an artefact of the chosen instantaneous excitation and not expected to match the experiment. The decay time of the anisotropy scales with the system size, in agreement with the onset of a macroscopic rotation of the simulated nanoparticle (faster for smaller crystals). In all traces, the anisotropy level shortly after time zero is the same and its value matches the experiment. The results remain slightly below the analytical estimations because the realistic atomic trajectories are irregular and not simple circles. A detailed analysis and more data are reported in the methods section.

The similarity of our numerical results with the measured Bragg spot anisotropies and the agreement of both results with a simple analytical model substantiate our interpretation of the measurement results as an ultrafast generation of circularly polarized phonons that carry almost all of the spin angular momentum that is missing from the demagnetization. Quantitatively, the main uncertainty in the analytical estimations is the average phonon frequency [37] and the main uncertainty in the experiment is the particular degree of demagnetization (see Fig. S14). Without any free parameter in the simulations, the absolute values of our three results are therefore consistent to each other within an estimated maximum mismatch of 30%. The sign of the anisotropy and the orientation of the polarization plane (see Fig. 2) are consistent for all three cases as well. Experimental evidences in the time domain provide further support: The anisotropic signature of the rotational phonons is generated quickly (see Fig. 2b) and maintains its alignment to the axis of the initial magnetization independently of a gradual increase of general lattice temperature (see Fig. 2a). Fitting to the hypothesis, the anisotropic lattice motion is therefore robust, appears as quickly as needed, consists of high-frequency phonons that align to the magnetic field normal and has an atomic mean square displacement that matches to the missing angular momentum.

**Consequences and relations to previous results**

In the simulations, the total energy associated with the phonon angular momentum is ~4 meV per nickel atom, which is several times smaller than the total energy that is deposited by laser absorption, roughly 25 meV per nickel atom. Only a fraction of the absorbed laser energy therefore contributes to the rotational phonons, while the rest simply heats the material via the remaining degrees of freedom. Consequently, the laser-induced excitation of the polarized phonons (causing/enabling demagnetization) competes with conventional electron-phonon coupling (heating the material), and the ratio of the two time constants controls the achievable degree of demagnetization. This finding explains why more demagnetization can occur in nickel at elevated base temperature [15] or why less and slower dynamics is measured for different chemical elements [2,4-6] with their distinct phonon dynamics. The reported results therefore link the success of phenomenological multi-temperature models [1,20,25,26,41] to an atomistic description.

The observed lattice disorder implies a localization of the excitation and break of translational crystal symmetry. Although the laser photons are coherent, excitation is quickly trapped in localized valence bands [21] or dislocations, and individual atoms (or at least very tiny domains) can therefore obtain local forces and a local angular momentum. If all atoms in the laser focus would be subject to identical forces, only macroscopic waves could emerge at the speed of sound, contrary to our observations of ultrafast anisotropic high-frequency disorder. Atomic-scale symmetry-breaking, fluctuations and non-concerted forces are therefore necessary to explain the measured diffraction results. The recent observation of a direct spin-strain coupling in iron [22] is in our picture a preferential coupling of spins to circularly polarized phonons near the center of the Brillouin zone and may be compatible with our results if we consider a potentially different degree of localization in this material [42].

Assorted experiments have shown that ultrafast demagnetization needs a finite time to occur, at least 150-300 fs in nickel [1,2,15,19,21,27] and several picoseconds in rare-earth metals [4-6]. Our results explain this bottleneck by the finite time that it takes to accelerate the polarized phonons in order to get rid of the spin angular momentum. The duration of this process depends on the amount of angular momentum in the spin system, the spin-orbit coupling, the strength of the coupling between spins and lattice, and the ability of the lattice to build up internal angular momentum. The lattice properties of a material and its susceptibility to carry polarized phonons therefore affect the speed and efficiency of the demagnetization. Note that electron diffraction becomes only sensitive to polarized phonons after the atoms have actually moved while demagnetization only requires their acceleration. The time constants of Fig. 2b are therefore upper limits for the time in which the angular momentum arrives in the lattice.

Once excited, the anisotropic phonons persist for picoseconds in the experiment and for tens of picoseconds in the simulations. We suppose that circularly polarized phonons are not free to thermalize into an isotropic equilibrium that obeys equipartition due to the necessary conservation of angular momentum. If so, the total angular momentum in a lattice can persist longer than the specific subset of high-frequency phonons that is initially excited. Macroscopic removal of angular momentum can only occur through low-frequency sound and strain waves that carry angular momentum away from the probing region and eventually mediate the onset of a macroscopic rotation [35]. Consequently, the traditional explanation of the Einstein-de Hass effect as a direct transfer of spin angular momentum to specimen rotation [22] should be amended to include an intermediate situation in which the angular momentum is carried by a rotational dynamics of the lattice [34-36]. In turn, a dedicated circularly polarized lattice excitation may produce an atomistic version of the Barnett effect and create magnetic fields [43-45]. The unifying principles are the conservations of energy and angular momentum on atomistic dimensions.

From an experimental perspective, the results show for the first time that ultrafast few-electron diffraction with terahertz-compressed pulses is a useful method for revealing structural dynamics in complex materials without any complications from space charge effects, neither regarding beam emittance nor pulse duration. The reported unification of single-electron pulses with all-optical control therefore offers an advantageous new tool for visualizing light-driven atomic motion in space and time.

To conclude, ultrafast electron diffraction of laser-demagnetized nickel reveals an almost instantaneous, long-lasting, non-equilibrium population of anisotropic phonons that oscillate predominantly in a plane perpendicular to the direction of the initial magnetization. A direct, efficient and ultrafast interaction of spins with high-frequency phonons is therefore decisive for the dynamics of magnetic materials. Theory and simulations links the observations to rotational lattice motion on atomic dimensions that takes up the missing angular momentum of ultrafast demagnetization before the onset of a macroscopic Einstein-de Haas rotation (see Fig. 4). Hence, circularly polarized and chiral phonons are not only a peculiarity of crystals with broken inversion symmetry [36,46-48] but can also be created in substantial amounts on femtosecond time scales with help of simple magnetic materials. This result suggests the possibility of utilizing phonon-based or phonon-assisted spin transport for applications in spintronics and for improving the speed and efficiency of ultrafast magnetic switching. A dedicated excitation and subsequent probing of polarized phonons or related hybrid quasiparticles with angular momentum may therefore provide innovative ways for understanding and optimizing the functionality of complex materials from an atomistic perspective.

**Acknowledgements:** We thank I. Wimmer for magnetic hysteresis data, B.-H. Chen for help with the optics, S. Geprägs for access to his x-ray diffractometer and F. Krausz for laboratory infrastructure. This research was supported by the European Union's Horizon 2020 research and innovation program via CoG 647771 and by the German Research Foundation (DFG) via SFB 1432.

**Author contributions:** PB and UN conceived the experiment. ST, MV and DE performed the diffraction experiments and analyzed the data. AB and ST produced the specimen under supervision of WK. AB and WK characterized the epitaxial growth. DK performed and analyzed the ultrafast optical measurements. UN conceived the theory and ME, HL and AD performed the simulations. PB, UN and ST wrote the manuscript with help of all coauthors.

**Competing interests:** The authors declare no competing interests.

**Data availability:** The data supporting the findings of this study are available from the corresponding author upon reasonable request.

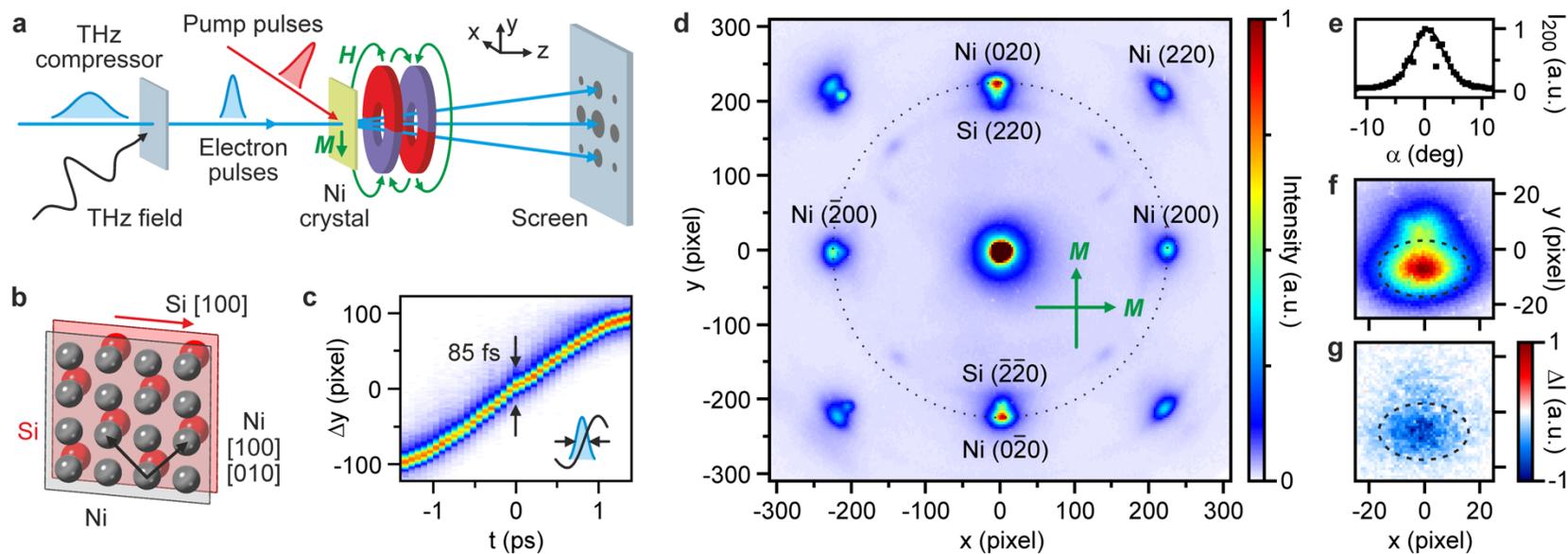

**Fig. 1.** Pump-probe electron diffraction measurements of atomic motions during ultrafast demagnetization. **a,** Experimental arrangement. Ultrashort electron pulses (blue) are compressed by THz radiation (black) and probe the laser-excited lattice dynamics in a single crystal of nickel (light green). Permanent magnets (blue and red poles) provide an initial magnetization (green) along distinct crystallographic axes. **b,** Epitaxial growth of nickel on silicon at 45°. **c,** THz streaking data for characterizing the all-optical compression of the electron pulses. Inset, schematics of the streaking of an electron pulse (blue) by a single THz fields cycle (black) for mapping time to position. **d,** Diffraction pattern before laser excitation. The green arrows denote different magnetic field directions that are probed. **e,** Rocking curve of the Ni($0\bar{2}0$) spot. **f,** Zoom into the Ni($0\bar{2}0$) spot (dashed area) and surroundings. **g,** Difference image around the Ni($0\bar{2}0$) spot; data before time zero is subtracted from averaged data between 2 and 4 ps.

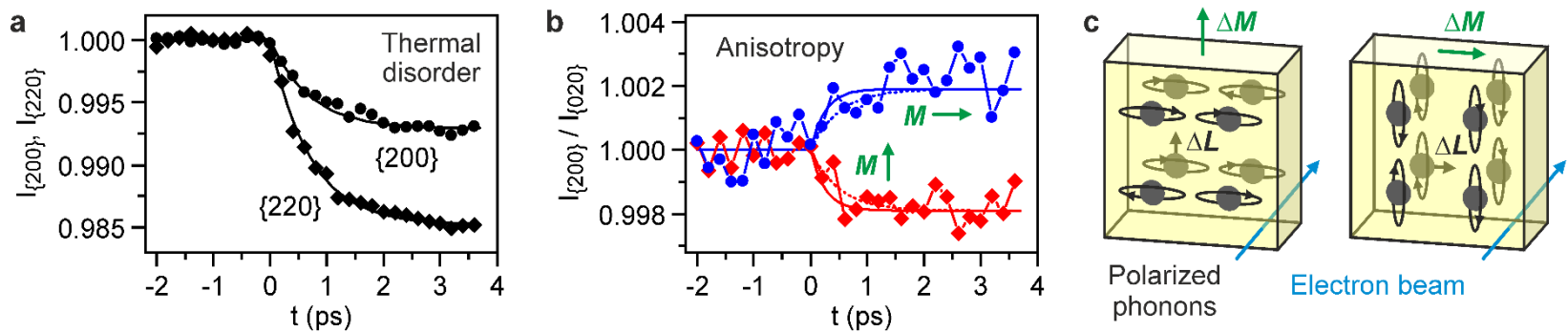

**Fig. 2.** Time-resolved diffraction results. **a,** Bragg spot intensities for two different diffraction orders as a function of time. The solid lines denote a decay time of 700 fs. Origin of this observation is electron-phonon coupling and thermal disorder. The standard error is $4\times10^{-4}$. **b,** Anisotropy between crystallographically equivalent Bragg spots of different orientation with respect to the initial magnetic field. $I_{\{200\}} = \frac{1}{2}(I_{200} + I_{\bar{2}00})$ and $I_{\{020\}} = \frac{1}{2}(I_{020} + I_{0\bar{2}0})$. The green arrows denote an initial magnetization along Ni[100] (blue data) or Ni[010] (red data). The standard error is $7\times10^{-4}$. The solid and dashed lines depict a rate constant of 200 fs and 500 fs, respectively. Data is equally compatible with both time constants. The measurement levels after saturation (1 ps < t < 4 ps) are 0.9981±0.0005 (red data) and 1.0019±0.0007 (blue data) under the constraint of an antisymmetric effect. **c,** Suggested origin of the results. A loss $\Delta M$ of magnetization implies a loss of angular momentum $\Delta L$ that is carried by atomic rotations. These rotationally polarized phonons locate in a plane that is perpendicular to the direction of magnetization. Electron diffraction (blue arrow) therefore reveals atomic disorder in an anisotropic way.

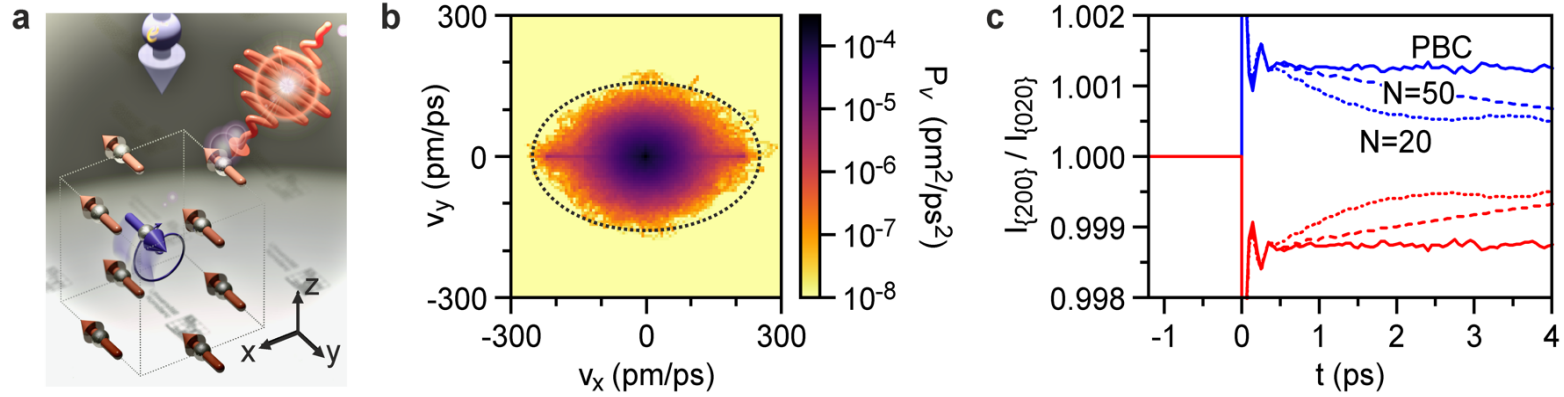

**Fig. 3.** Molecular dynamics simulations. **a,** After laser excitation (red pulse), we excite the lattice by instantaneously displacing an atom locally from its equilibrium position assuming a velocity in the *x-z*-plane perpendicular to the displacement (blue motion), leading to an angular momentum of $\hbar$ in *y* direction (blue), compensating the lost spin angular momentum (blue arrow). Bragg peaks are calculated for the indicated electron beam direction (light blue). **b,** Nickel atoms develop an anisotropic velocity distribution, violating equipartition. Motion is faster (and displacements are larger) in *x*-direction than along *y*. **c,** Simulated electron diffraction and Bragg spot anisotropy in analogy to Fig. 2b. Blue and red data denote an initial magnetization along Ni[100] and Ni[010], respectively. In our simulations this corresponds to swapping the definition of *x* and *y* axes, so that our data are simply mirrored. N=20 for about $4\times20^3$ nickel atoms; N=50 for about $4\times50^3$ nickel atoms, both with open boundary conditions; PBC for about $4\times20^3$ nickel atoms with periodic boundary conditions. The initial anisotropy level after laser excitation is similar in all cases and fits to the experiment (see Fig. 2b).

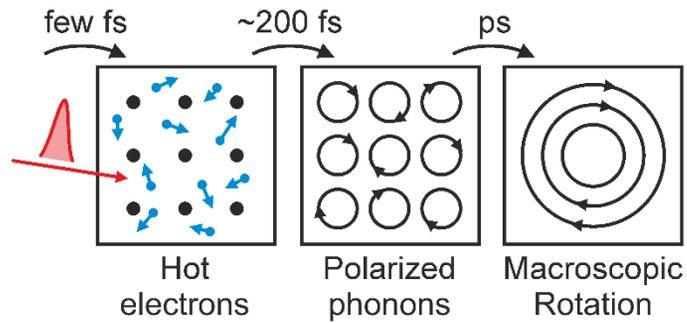

**Fig. 4.** Sequence of events; compare Ref. [35]. After the laser (red) has created hot electrons with spin (blue), they quickly couple to anisotropic and circularly polarized phonons with angular momentum (middle panel). The material is demagnetized at this point. Only later, low-frequency shear waves mediate a macroscopic rotation of the specimen as a whole (right panel).

# Methods

**Thin film growth**

A 22 nm thick Ni layer was deposited on a 0.1 mm thick and 3 × 3 mm² large single-crystalline Si(100) transmission electron microscopy (TEM) grid, henceforth called the substrate. The TEM grid (Plano Inc.) exhibits eight 100 × 100 µm² large silicon membranes plus one marker membrane with 35 nm thickness in a 3×3 arrangement. Prior to deposition, the TEM grids are ultrasonically cleaned using consecutive baths of acetone and ethanol, followed by a rinse with deionized water. Subsequently, the substrate is dipped in 5 % hydrofluoric acid for approximately 15 s, after which it is blown dry by laboratory grade clean air and immediately inserted into the thin film deposition system. This treatment removes the native silicon oxide layer and prepares a hydrogen-terminated silicon surface for epitaxial growth [1-4].

Thin film preparation follows the technique of metal-metal epitaxy on silicon (MMES) [5-7]. We use direct current (DC) magnetron sputtering, carried out in our thin film deposition system as described in Refs. [8,9]. One of the three 2" sputter guns is loaded with a 6.3 mm thick, 99.99 % pure Cu sputter target and a second sputter gun is loaded with a 1.4 mm thick 99.99 % pure Ni target. The distance between the sputter target and the substrate is approximately 11 cm normal to the surface of the substrate. The film thickness is time-controlled with a deposition shutter that is located between the sputtering target and the substrate.

The system is pumped down to a base pressure of approx. $3\times10^{-8}$ mbar and the substrate is cooled to 273 K, where it is allowed to stabilize for 1 h before commencing the deposition. To remove any potential dirt from the sputtering target, we apply 10 minutes of pre-sputtering with closed deposition shutter. An initial 6 nm of Cu seed layer is deposited at an ultra-high purity (7N) Ar working gas pressure of $2.55\times10^{-3}$ mbar and a sputtering power of 50 W, resulting in a deposition rate of 0.5 Å/s. Consecutively, the Ni layer is deposited onto the seed layer at a working gas pressure of $4.0\times10^{-3}$ mbar and a DC sputtering power of 50 W. These parameters result in a deposition rate of 0.41 Å/s.

*X-ray analysis of the sample.* After deposition, the sample is characterized ex-situ by x-ray reflectometry (XRR) as well as out-of-plane (oop) and in-plane (ip) x-ray diffraction (XRD). In the following, $\lambda$ is the wavelength of the applied x-ray radiation, $B(2\theta)$ is the full-width-at-half-maximum (FWHM) of the diffraction peak and $\theta$ is the angular peak position. Supplemental $\omega$-scans (also known as rocking-scans) are performed for the XRD case, which allows to determine the mosaic spread, viz. the directional alignment of the crystallites. For a Gaussian shaped rocking-scan, the mosaic spread is quantified by its corresponding standard deviation $\sigma = \frac{\text{FWHM}}{2\sqrt{2ln2}}$ [10]. The size of the crystallites is analogous to the coherence length $L$, which can be estimated by the Scherrer formula [11] $B(2\theta) = \frac{0.94\lambda}{L\cos\theta}$ with $B(2\theta)$ and $\theta$ in radians units.

*Instrumentation.* The oop structural analysis was performed using a two-circle x-ray diffractometer with parallel beam optics and $\text{Cu}_{K\alpha}$ source in a $\theta - 2\theta$ scattering geometry (D5000, Siemens GmbH). The instrument resolution of the x-ray diffractometer is 0.023°, given by the FWHM of the Si(004) out-of-plane rocking-scan. This value is small enough such that instrumental broadening in the analysis of the deposited Cu and Ni films can be neglected. For ip structural analysis, we apply a four-circle x-ray diffractometer, also with parallel beam optics and $\text{Cu}_{K\alpha}$ source in a $\theta - 2\theta$ scattering geometry (D8, Bruker GmbH). It offers additional axes for sample inclination ($\chi$) relative to the scattering plane and axis rotation ($\phi$) around the sample normal. This capability allows to access ip crystal information. An instrument resolution of 0.031° is determined from the FWHM of the Si(004) out-of-plane rocking-scan.

*X-ray reflectivity.* For XRR analysis, the sample is aligned with respect to its surface by maximizing the intensity in the regime of total reflection. Quantitative analysis of the XRR data was carried out using the GenX 3.0.0 software package [12]. For data fitting, the logarithmic figure of merit FOM $\sim \sum |logR_{\text{fit}} - logR_{\text{meas}}|$ is used, where $R_{\text{fit}}$ is the fitted reflectivity and $R_{\text{meas}}$ is the measured reflectivity, respectively. We assume a layer structure of $NiO_x$ on Ni on Cu on Si. $NiO_x$ denotes the natural nickel oxide layer that is expected to form at the surface after the thin film deposition process and a small additional layer of unknown surface coverage (dirt) is also allowed in the fits. Figure S1a depicts the measured x-ray reflectivity data and table 1 shows the best fit results. The scattering length density profile is shown in Fig. S1b.

| Layer | Thickness [Å] | Density [g/cm³] | rms roughness [Å] |
|-------|---------------|-----------------|-------------------|
| Dirt  | $1 \pm 2.5$   | N. A.           | $7.7 \pm 0.2$     |
| $NiO_x$ | $31 \pm 1.9$ | $5.11 \pm 0.18$ | $17.4 \pm 3.6$    |
| Ni    | $222 \pm 3.0$ | $9.32 \pm 0.26$ | $6.4 \pm 1.3$     |
| Cu    | $63 \pm 2.1$  | $8.43 \pm 0.15$ | $7.7 \pm 1.3$     |
| Si    | $\infty$      | $2.32 \pm 0.01$ | $3.5 \pm 0.1$     |

**Table 1.** The best fit parameters obtained by fitting the XRR intensities, assuming a four-layer model. The errors are estimated by a 5 % increase over the optimum logarithmic figure of merit.

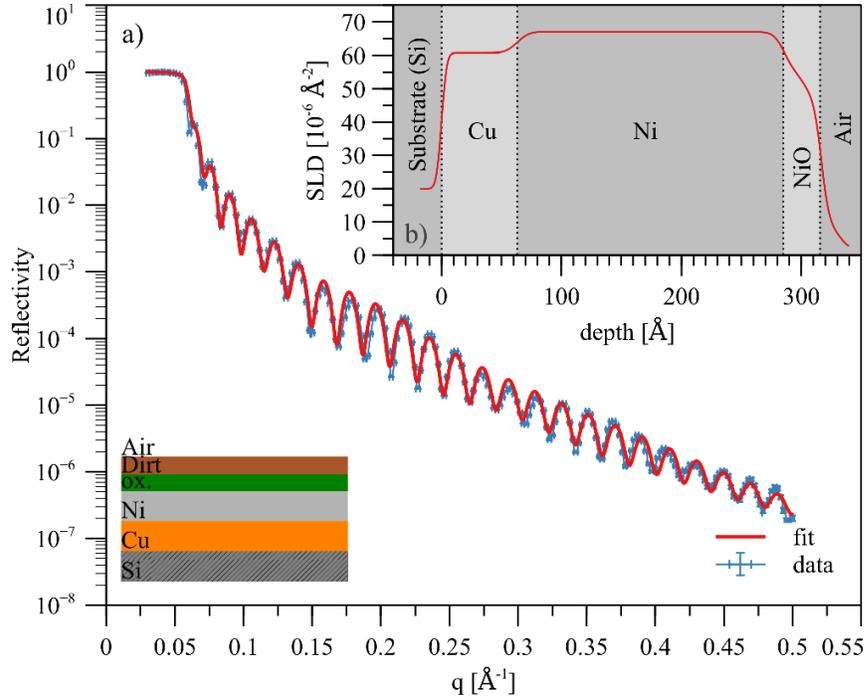

Fig. S1. X-ray reflectivity data and fit of the sample using a four-layer model. (a) Data and fit. (b) scattering length density profile used for the fitting. The dashed lines indicate the slab model of the corresponding layers. The inset shows the layer sequence.

*Out-of-plane X-ray Diffraction.* For XRD analysis, the sample is aligned with respect to the Si(004) oop reflection as a well-defined reference for the determination of the crystalline directions of the deposited Cu and Ni films. In the following, $a$, $b$ and $c$ represent the lattice constants along the crystallographic directions [100], [010] and [001]. Fig. S2 depicts the oop XRD measurements in the range $40° \leqslant 2\theta \leqslant 60°$. The observed intensities at $2\theta \sim 50.53°$ and $2\theta \sim 52.13°$ correspond to the Cu(002) and Ni(002) reflections with c-axis lattice constants of $c_{Cu} = 3.612\pm0.003$ Å and $c_{Ni} = 3.509\pm0.003$ Å, respectively.

By applying the Scherrer formula (see above), we estimate a coherence length of $L(001)_{Cu} = 159.5\pm2.7$ Å and $L(001)_{Ni} = 221.1\pm2.7$ Å. These lengths are of the same order of magnitude as the corresponding layer thicknesses. This result indicates that there is no break of the epitaxy in growth direction. Also, the rocking-scans (see inset of Fig. S2) reveal no indication of any crystal defects along the growth direction. In particular, we observe no splitting of the rocking-curve into multiple peaks [4]; the measured FWHM values of $FWHM_{Cu} = 5.6°\pm0.1°$ ($\sigma_{Cu} = 2.378\pm0.042$) and $FWHM_{Ni} = 4.4°\pm0.1°$ ($\sigma_{Cu} = 2.378\pm0.042$) are in agreement with literature values [4,5,10] for Cu and Ni films of similar thickness. Note that the electron rocking curve (see Fig. 1e) is slightly wider due to electron energy losses.

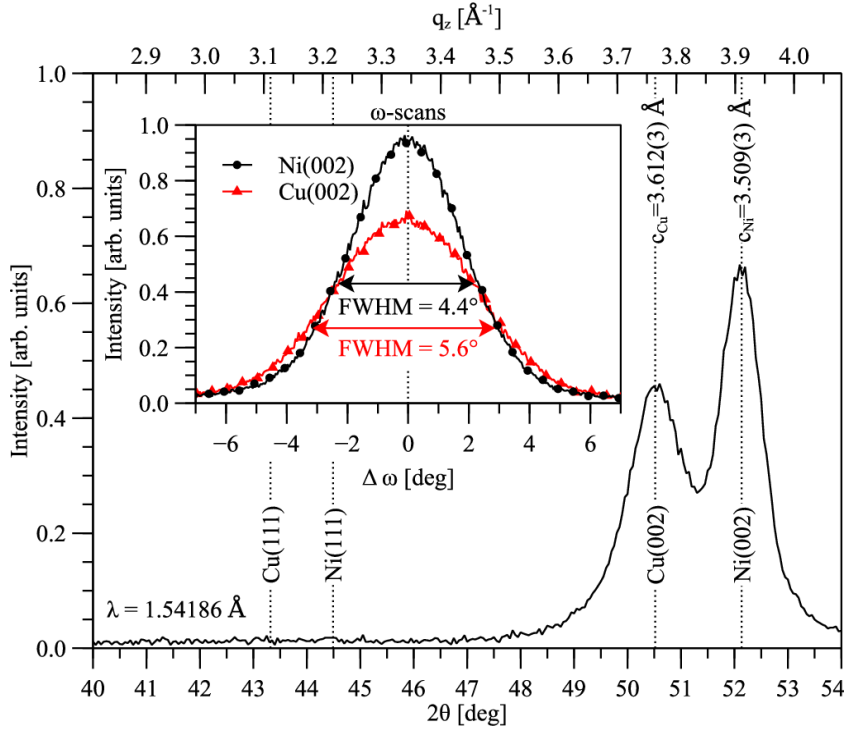

**Fig. S2.** Out-of-plane XRD scan in the angular regime of $40° \leqslant 2\theta \leqslant 60°$. The observed intensities at $2\theta \sim 50.53°$ and $2\theta \sim 52.13°$ correspond to Cu(002) and Ni(002). The lack of any Cu(111) and Ni(111) intensities confirms the epitaxial growth. The inset graph shows the rocking-scans over the Cu(002) and Ni(002) peak positions.

*In-plane X-Ray diffraction.* An oop analysis alone is not sufficient for proof of an epitaxial growth, because crystalline misalignment can also occur ip. Additionally, we aim for determining the crystallographic orientation of the Cu and Ni layers with respect to the substrate. Therefore, ip sensitive scans are performed with the sample being aligned in $\chi$ and $\phi$ with reference to the Si(004) and Si(111) substrate reflections. Setting the $2\theta$ angle to the Si(111), Cu(111) and Ni(111) ip peak positions and rotating the sample around its normal $\phi$ at the corresponding inclination angle $\Delta\chi = 54.74°$ between the (001) oop and (111) ip crystal directions provides information on the ip crystal symmetry. The three $\phi$-scans for the Si(111), Cu(111) and Ni(111) ip peaks are shown in Fig. S3.

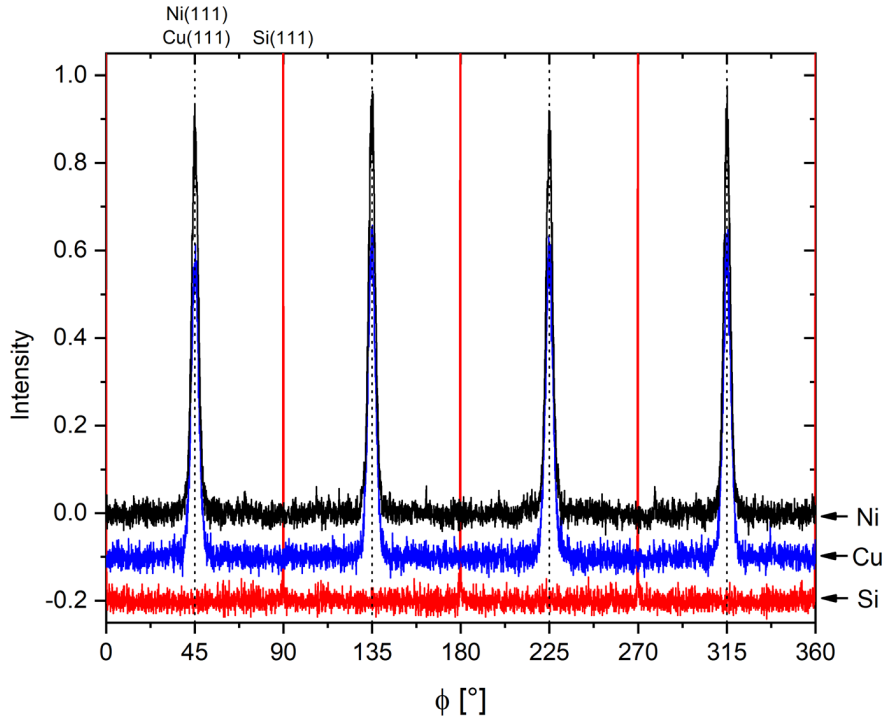

**Fig. S3.** $\phi$-scans for the Ni(111), Cu(111) and Si(111) ip peaks, obtained at an inclination angle of $\Delta\chi = 54.74°$. A clear fourfold symmetry of the Cu(111) and Ni(111) ip reflections is observed with an offset angle of 45° to the Si(111) substrate reflections. For reasons of clarity, the scans are shifted in intensity by -0.1 each.

A clear fourfold symmetry of the Cu(111) and Ni(111) ip reflections is observed with an offset angle of 45° to the Si(111) substrate reflections. This result demonstrates that Ni grows epitaxially on the Si(100) surface with the relationship Ni[001] ∥ Cu[001] ∥ Si[110] and Cu(001) ∥ Ni(001) ∥ Si(001). The observed intensities at $2\theta \sim 43.41°$ and $2\theta \sim 44.47°$ (see Fig. S4) correspond to the Cu(111) and Ni(111) reflections with diffraction plane spacings of $d_{Cu}(111) = 2.084 \pm 0.004$ Å and $d_{Ni}(111) = 2.037 \pm 0.003$ Å, respectively. Assuming $a = b$, the ip lattice parameters of the Cu and Ni-films are obtained via geometrical relationships as $a_{Cu} = b_{Cu} = 3.608 \pm 0.008$ Å and $a_{Ni} = b_{Ni} = 3.539 \pm 0.008$ Å. In [111] direction, we obtain coherence length values of $L(111)_{Cu} = 258.8 \pm 1.8$ Å and $L(111)_{Ni} = 305.7 \pm 1.7$ Å, from which we infer lateral coherences of $L(100)_{Cu} = L(010)_{Cu} = 144.1 \pm 2.1$ Å and $L(100)_{Ni} = L(010)_{Ni} = 149.3 \pm 2.6$ Å. These values are comparable to the coherence length of the electron beam (~200 Å).

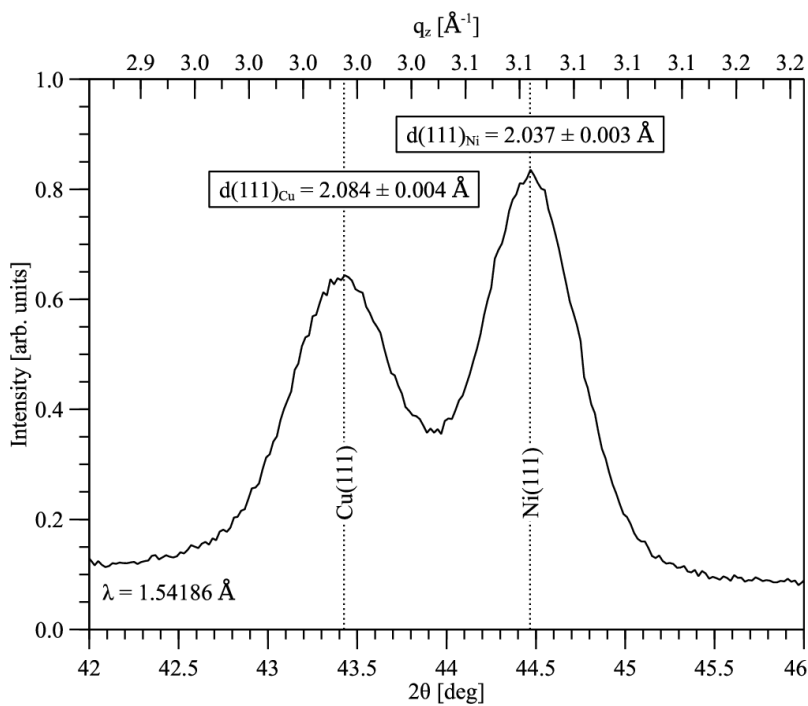

**Fig. S4.** In-plane XRD scan at an inclination angle $\Delta\chi = 54.51°$. The intensities at $2\theta \sim 43.41°$ and $2\theta \sim 44.47°$ correspond to the Cu(111) and Ni(111) reflections, respectively.

The observation of weak Cu(111) and Ni(111) reflections for $\Delta\chi = 79.00°$ at $\phi \sim 10°$, 80°, 100°, 170°, 190°, 260° and 350° (red curve in Fig. S5) and for $\Delta\chi = 15.80°$ at $\phi \sim 45°$, 135°, 225° and 315° (blue curve in Fig. S5) indicate the presence of occasional (111)-plane twin defects [13,14].

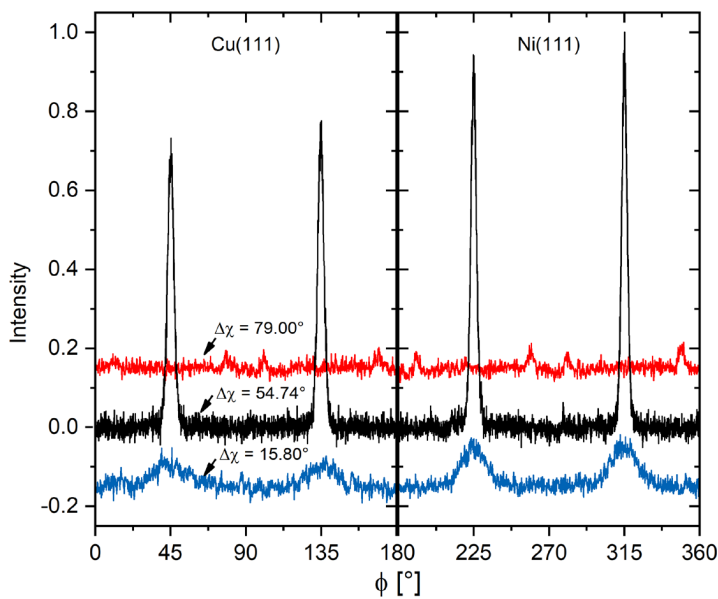

**Fig. S5.** $\phi$-scans for the Cu(111) and Ni(111) reflections, obtained at inclination angles of $\Delta\chi = 15.80°$, $\Delta\chi = 54.74°$ and $\Delta\chi = 79.00°$. For clarity, the scans are shifted in intensity by ±0.15. Cu(111) intensities are shown in the angular regime of $0° \leq \phi \leq 180°$, while the Ni(111) intensities are shown for $180° \leq \phi \leq 360°$.

For the Ni(111) reflections, the strongest intensity is observed at an inclination angle of $\Delta\chi = 54.51°$ instead of $\Delta\chi = 54.74°$ which would be expected for a cubic system. This mismatch indicates a slight tetragonal distortion of the Ni film, confirmed by ip diffraction scans in the range $42° \leqslant 2\theta \leqslant 46°$ at an inclination angle of $\Delta\chi = 54.51°$ (see Fig. S4). However, such a small distortion is irrelevant for the ultrafast electron diffraction experiments, because the electron beam divergence and rocking curve widths far surpass the measured angle differences.

**Magnetic hysteresis measurements and design of the permanent magnets**

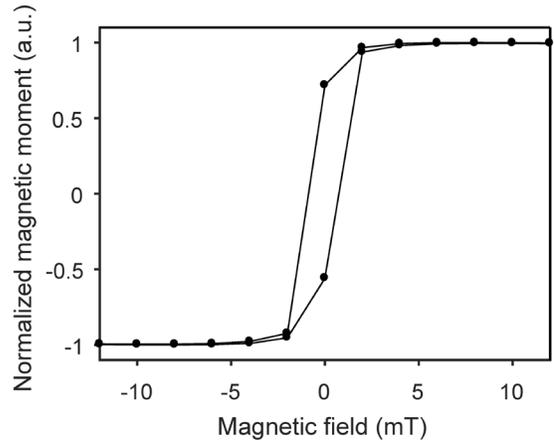

The magnetic hysteresis curve of our nickel specimen and the field needed for sufficient in-plane magnetization by the permanent magnets are determined with a superconducting quantum interference device (MPMS XL5 SQUID, Quantum Design). Figure S6 depicts the results. The data indicates saturation at ~4 mT. The permanent-magnetic structure for defining and maintaining a constant in-plane magnetization within the sample is a construction of two NdFeB ring magnets with 10 mm outer diameter, 7 mm inner diameter, 3 mm thickness and a remanent magnetization of ~1.3 T. Two diametrically magnetized ring magnets are aligned with opposite poles facing each other, providing at the specimen a magnetizing field of ~10 mT in the direction along the surface (see Fig. 1a), more than two times larger than the field required for saturation. Boundary element simulations (CST Studio suite, Dassault Systèmes) show that even with minor angular misalignments of up to 15° or residual sideways displacement of up to 100 μm the effective magnetic field at the sample is largely homogeneous and oriented in-plane.

**Fig. S6.** Magnetic hysteresis curve of our nickel specimen, obtained by an in-plane SQUID measurement.

**Laser fluence determination and optical absorption depth**

The optical beam radius $\omega_0 \approx 150$ μm (at an intensity of $1/e^2$) at the sample position is evaluated by knife-edge measurements. Using the measured average laser power $P_{avg}$ within the vacuum chamber and the laser repetition rate $f_{rep}$, we obtain the excitation fluence via $F_{peak} = \frac{1}{\cos(25°)} \frac{2}{\pi \omega_0^2} \frac{P_{avg}}{f_{rep}}$. The cosine term accounts for the stretching of the horizontal beam profile at 25° laser incidence. The optical absorption of our multilayer specimen as a function of depth is calculated with finite-time-domain difference simulations. We use Bloch boundary conditions and an incidence angle of the light wave of 25°. Figure S7 depicts the results. The upper panel shows the real and imaginary part of the refractive index of the layer materials (dotted and dashed, respectively) and the obtained electric field amplitude $E$ in reference to the incoming amplitude $E_0$ (solid line). The lower panel shows the simulated optical absorption density for the parameters of the experiment. We see an almost constant profile throughout all of the nickel layer (green). The simulated reflectivity is ~60 % and the transmission is ~10 %. More than 95 % of the deposited pulse energy goes into nickel and less than 5 % into the copper seed layer. Although the silicon substrate does not absorb laser light, it might still heat up a little by hot electron transport, because the difference between the work function of nickel (~5 eV) and the electron affinity of silicon (~4 eV) is comparable to the silicon bandgap (1.1 eV).

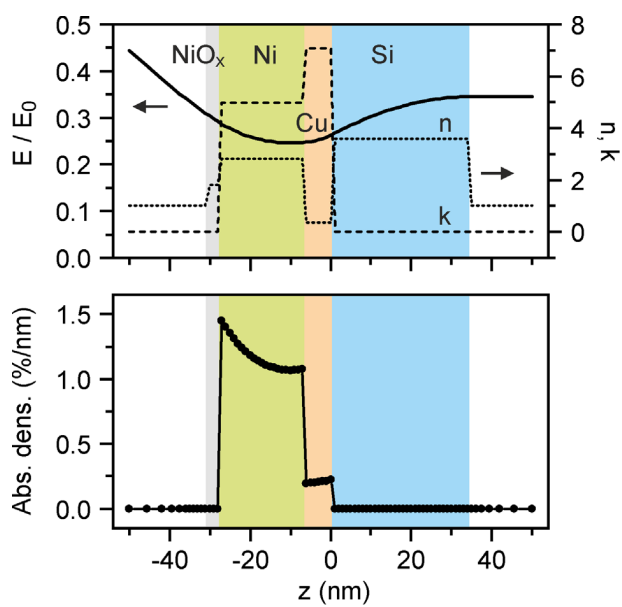

**Fig. S7.** Simulated optical energy disposition as a function of penetration depth. Upper panel: solid line, normalized electric field amplitude; dotted line, real part of the refractive index; dashed line, imaginary part of the refractive index. The laser comes from the left. Lower panel: absorption as a function of depth. The green, red, blue and grey areas denote nickel, copper, silicon and NiO$_x$, respectively.

**Base temperature of the pump-probe experiment**

Important for pump-probe experiments is a back-cooling of the specimen between the excitations. The base temperature of the sample before laser excitation is determined by comparing Bragg diffraction before time zero with diffraction pattern taken without any laser excitation at all. This change is less than 0.5 % for the Ni(200) spot and less than 1 % for the Ni(220) spot. Using reported values for the mean square displacement $\overline{u^2}$ as a function of temperature [15], we calculate the Debye-Waller-Factor $\frac{I}{I_0} = exp\left(\frac{-1}{3}|G|^2\overline{u^2}\right)$ as a function of temperature; $G$ is the reciprocal lattice vector. We infer that the specimen has a base temperature not larger than ~50 K above the laboratory temperature before each excitation with the pump laser. Alternatively, we consider the average laser power of ~35 mW of which ~15 mW are absorbed. A numerical heat conduction model with nickel's heat conductivity of ~1.3×10$^{-5}$ m$^2$/s, the laser spot as a source and a heart sink given by the macroscopic sample edges reveals a base temperature increase of ~50 K. Therefore, no substantial heat accumulates on average over the course of the experiment.

**Bragg spot analysis procedure**

For analysis of time-dependent Ni Bragg spot intensities, an elliptical region is selected for each Bragg spot in a way that minimizes overlap with the adjacent Si peaks. For each image and Bragg spot, intensity in this region is summed up. About 250 pump-probe scans over the measured time range are performed rapidly in order to average out potential long-term drifts of the experiment. In Fig. 2b, the data is scaled to an anisotropy of one at delays below zero. Averaging the time evolution of Friedel pairs (opposing Bragg spots) makes the experiment particularly insensitive to potential time-dependent deviations of the Bragg condition. In addition, the rocking curve is broad (see Fig. 1e). We therefore probe exclusively sub-unit-cell dynamics and atomistic disorder, not mechanical strain waves or torques.

# Electron rocking curves and multiple scattering effects

The angle-dependent intensity of each Bragg peak within the diffraction pattern is recorded with the ultrafast electron diffraction beamline at the same experimental conditions and beam parameters as in the pump-probe experiments. Only the terahertz compressor is turned off and the Al foil is removed for simplicity. The sample is rotated ±20° around the [010] axis with 0.25° steps and rotated by ±2.5° around the [100] axis, limited by the mechanical range of our goniometer. During the scan, the sample's spatial position is continuously corrected to keep the specimen central to the electron beam. The data in Fig. S8 shows the rocking curve of the Ni(200) peak. The full width at half maximum is 6.5°. The narrow spike at ~1° and ~3° are attributed to the effects of the silicon substrate. In the pump-probe experiment, such special angles are mostly avoided (see Fig. 1f). The slight strain of the growth (see above) is helpful for this purpose.

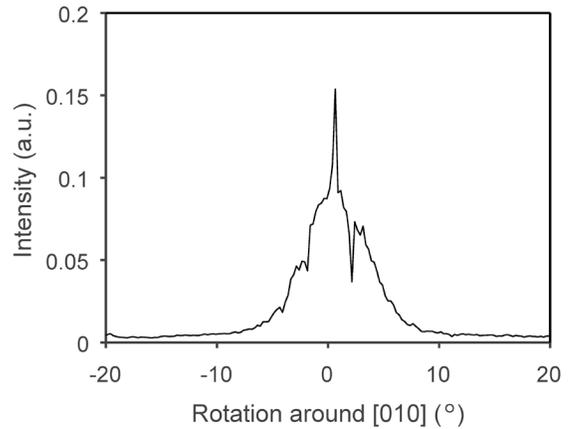

**Fig. S8.** Rocking-scan data obtained with the femtosecond electron beam. Shown is the Ni(200) peak when rotating the specimen around the [010]-axis.

# Laser pulse compression

In order to reduce the optical pulse duration from our Yb:YAG laser system (270 fs), we apply a series of two self-defocussing bulk media [16]. First, we use an $f = 75$ mm lens and an 8-mm beta barium borate (BBO) crystal (Döhrer Electrooptic GmbH). The broadened spectrum is compressed with a grating sequence (LSFSG-1000-3212-94, LightSmyth) at a distance of 2.5 mm. Second, we apply an $f = 40$ mm lens and a 6-mm BBO crystal. The output pulses are compressed with a 12 mm thick block of high-refractive-index glass (SF10) with a group delay dispersion of ~1300 $fs^2$. Residual radiation at 515 nm wavelength is filtered out by high-reflection mirrors. Frequency-resolved optical gating (FROG) is used to characterize the pulse duration (see Fig. S9), revealing a full width at half maximum of 90 fs. A replica of the air/vacuum entrance window in front of the FROG setup ensures a correct consideration of dispersion.

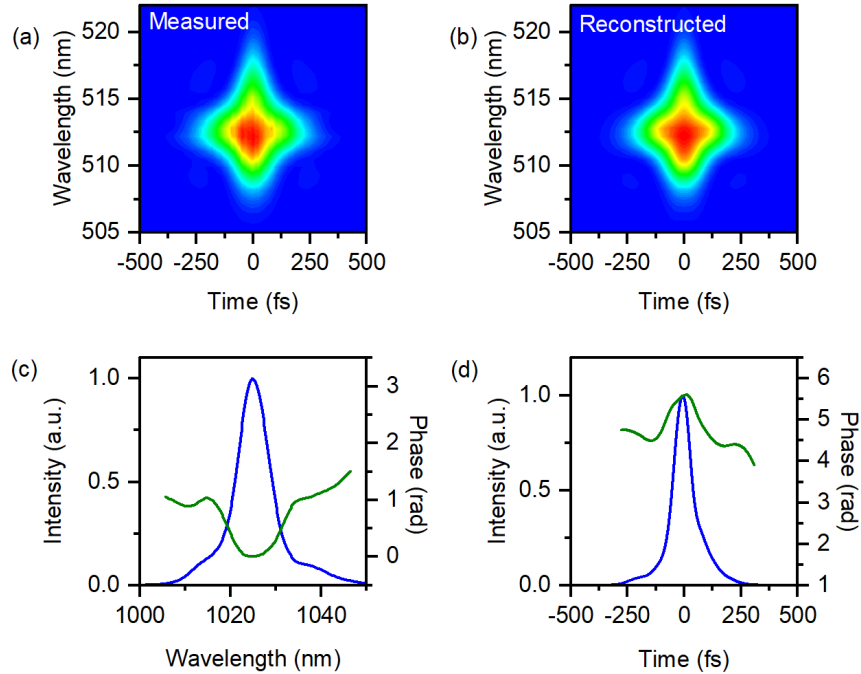

**Fig. S9.** Second-harmonic-generation FROG measurements of the optical output pulses after compression. (a) Measured FROG trace. (b) Retrieved FROG trace at 0.3 % FROG error. (c) Evaluated spectrum (blue) and spectral phase (green). (d) Retrieved pulse shape (blue) with temporal phase (green). The pulse duration is 93 fs.

**Electron pulse generation and compression by terahertz single-cycle pulses**

Ultrashort electron pulses are generated by two-photon photoemission [17] and accelerated to a kinetic energy of 70 keV. We use only 10-20 electrons per pulse, just below the onset of detectable space-change effects. The electron beam diameter (full width at half maximum) at the specimen is ~60 µm. Despite the absence of space charge forces, the dispersion of free space for nonrelativistic electrons lengthens the pulse duration to ~500 fs at the specimen location [18,19]. To overcome this limit, we apply electron pulse compression by THz single-cycle pulses in planar geometry [18]. The electron beam crosses the THz pulses at an angle of approximately 90° at an ultrathin mirror made of a 10 nm Al layer on a 10 nm SiN support membrane and placed at a distance of 0.4 m from the electron source and 0.5 m before the specimen. All-optical streaking [19] is used to characterize the electron pulse duration directly at the specimen location (see Fig. 1c). The streaking element is a butterfly-shaped copper resonator (400 µm width, 400 µm height, 56 µm gap width, 80 µm gap height) that is excited by ~$10^6$ V/m of THz peak field strength. A 50-µm aperture improves the resolution of the streaking signal. Electron pulse durations of <130 fs, <120 fs, <75 fs and <80 fs (upper limits due to streaking resolution) are obtained for the diffraction scans that are combined in Figs. 2a-b. The time mismatch originating from the 25° laser-membrane angle is ~100 fs. The temporal apparatus function is ~150 fs (full width at half maximum). Recording meaningful electron diffraction images with single-electron pulses implies data accumulation over many pump-probe cycles. Measured diffraction patterns are therefore statistical averages over space and time [17].

**Molecular dynamics simulations**

Our molecular-dynamic simulations are carried out with LAMMPS [20,21] using the implemented symplectic Velocity-Verlet method which ensures energy conservation and, for open boundary conditions, angular momentum conservation. A nickel fcc lattice of size $N \times N \times N$ unit cells ($\approx 4N^3$ atoms) is simulated using the appropriate potential, which was calculated via the embedded-atom method (EAM) and is implemented in LAMMPS [22]. Starting from its ground state the lattice is excited accordingly by an initial condition as described below. To compare the resulting dynamics with the experimental measurements, time-dependent diffraction patterns $I(\vec{k}, t)$ are calculated using the selected area electron diffraction method (SAED) of LAMMPS [23] for electron diffraction with an incident radiation from direction $\vec{e}_z$ and a wave length $\lambda = 0.04635$Å. We simulate $N_{av} = 10, \ldots, 100$ different initial conditions (depending on the system size) and average the resulting intensities $\langle I \rangle_{N_{av}}$. The integrated peak intensities, from which the peak contrast is calculated, are defined by

$$I_{(YZ0)}(t) = \int_{|\vec{k} - \vec{k}_{(XY0)}| \leq \varepsilon} \langle I(\vec{k}, t) \rangle_{N_{av}} \, d^2k, \tag{S1.1}$$

where $\varepsilon = 0.1\text{Å}^{-1}$ is the radius in momentum space the diffraction pattern is integrated over. We study three different systems: (i) For a nickel lattice with open boundary conditions along all directions, energy- and angular momentum conservation is ensured. A global rotation of the crystal is possible and boundary effects as well as finite-size effects can be observed. The ground-state atomic positions have to be simulated before it is excited via the initial condition. (ii) For periodic boundary conditions along all directions only energy conservation is ensured. No global rotation can occur, no boundary effects can be observed, but finite size effects still exist. (iii) To investigate a global rotation of the system according to the Einstein-de Haas-effect, no direct molecular dynamics simulation is performed, but the system in its ground state is rotated by the angle $\varphi = |\vec{\omega}| \cdot t$, where $\vec{\omega} = \Theta^{-1} \cdot \vec{L}_0$ with the total angular momentum $\vec{L}_0$ and the inertia tensor $\Theta$.

*Initial condition.* We mimic an excitation of the lattice with finite angular momentum $\vec{L}_0$, which takes over the spin angular momentum corresponding to a demagnetization process of $\Delta \vec{M} = -\Delta M \cdot \vec{e}_y$ via appropriate initial conditions, suitable for molecular dynamics simulations. Within our model, the angular momentum is transferred instantaneously and locally to a fraction of atoms $\varrho$. Note that a sole velocity excitation, for instance, does not excite any atomic angular momentum, since any atomic motion which starts in an equilibrium position would have the displacement vector aligned with the velocity. Therefore, each excited atom is deflected from its equilibrium position by $\vec{\Delta} = \Delta(\cos(\varphi), 0, \sin(\varphi))$ and given a velocity $\vec{v}_0 = v_0(-\sin(\varphi), 0, \cos(\varphi))$, such that the local angular momentum equals $\hbar$. The angle $\varphi$ is uniformly distributed in $[0, 2\pi]$ under the side condition explained below. The resulting angular momentum shows in $-\vec{e}_y$-direction perpendicular to the excitation plane. Aiming for a local angular momentum of $m_{\text{Ni}}|\vec{\Delta} \times \vec{v}_0| = \hbar$, the excitation is chosen with minimal energy, i.e. with $\Delta$ being at the minimum of the effective potential.

We further aim for zero total velocity of the crystal, $\sum_l \dot{\vec{r}}^l = 0$ and that the total angular momentum exactly equals $\vec{L}_0$. This is not trivial for a system of finite size, such that we use side conditions: First, only half of the excited atoms are randomly selected. Second, symmetric partners of the randomly-selected atoms have to be chosen with respect to the crystal's center of mass. This ensures $m_{\text{Ni}} \sum_l \vec{r}^l \times \dot{\vec{r}}^l = \vec{L}_0$. And third, half of these symmetric pairs is excited by $\vec{\Delta}$ and $\vec{v}_0$ whereas the other half is excited in the opposite direction by $-\vec{\Delta}$ and $-\vec{v}_0$, which delivers zero total velocity. For an implementation in LAMMPS with periodic boundary conditions, our initial conditions require a slightly different algorithm, which also yields a fixed total angular momentum and zero total velocity.

The fraction of excited atoms $\varrho$ can be determined from the saturation magnetization of nickel $M_s = \mu_s/V_{\text{uc}} = 0.616\, \mu_B/V_{\text{uc}}$ [24] (where $V_{\text{uc}} = \frac{1}{4}a^3$ with $a = 352$ pm is the volume of an fcc nickel unit cell) and the demagnetization $\Delta M$ via $\Delta M = \left|\gamma \frac{\vec{L}_0}{V}\right| = |\gamma| \frac{\varrho \hbar}{V_{\text{uc}}}$, where $\gamma = g' \frac{\mu_B}{\hbar}$ with $g' = 1.8550$ [25]. A demagnetization of 27 %, 42 % and 63 %, which are the values investigated in this paper, correspond to $\varrho \approx 0.09, 0.14, 0.21$ (see also Fig. S14).

*Finite-size effect and comparison to periodic boundary conditions.* As the numerical investigation is limited to finite system sizes, we investigate the influences by testing $N = 10, 20, 50$ (about $4 \times 10^3$, $3.2 \times 10^4$ and $5 \times 10^5$ atoms) for open boundary conditions (OBC). Figure S10, left panel, depicts the anisotropic contrast $I_{\{200\}}/I_{\{020\}}$ resulting from our molecular dynamics simulations. All systems show a very rapid decay of the anisotropic contrast in $\approx 50$fs, a time scale which is not in the focus of our investigation, because it is shorter than the laser pulse that we do not model explicitly. However, we also find a slower decay on the time scale of some picoseconds, which is not observed in the experimental data (see Fig. 2b). This decay time scales towards a longer-living anisotropy with system size and is therefore a finite-size effect. Figure S1.1 shows simulation data for systems with periodic boundary conditions (PBC) which avoids surface effects. In such a system, the initial angular momentum $\vec{L}_0$ is not conserved but rather relaxes very quickly on a time scale of tens of femtoseconds. Nevertheless, the anisotropy remains substantial and preserves even longer than for open boundary conditions. Our interpretation is that the mode coupling—either of transversal and longitudinal modes or of transversal modes along different directions—causes the equilibration, a mechanism that appears less efficient for periodic boundary conditions due to the lack of phonon reflections at the boundaries. For open boundaries, the finite-size scaling follows from the fact that for larger systems surfaces are less important. Note that the contrast in the mean-squared velocities $2\langle v_y^2\rangle/(\langle v_x^2\rangle+\langle v_z^2\rangle)$ shows a very similar behavior, see right panel of Fig. S10, where the long-time evolution is depicted.

Furthermore, we can rule out a global Einstein-de Haas rotation as explanation for the observed contrast by means of a comparison with results from a global system rotation as discussed above, which assumes the total angular momentum $\vec{L}_0$ being instantly and completely transferred to a rigid-body rotation. The effect on diffraction requires tens of picoseconds to develop, evident from the data plotted in the right-hand side of Fig. S10. On the time scale of the experiment, rotation effects are negligible. Furthermore, the corresponding angular velocity in the simulation is prone to finite system size with $\omega \propto N^{-2}$, which means that it is even less relevant for the experimental system sizes.

In conclusion, the finite-size effects in our simulations are essentially boundary effects. This suggests that anisotropic peak intensities in diffraction could persist on time scales up to 100 ps.

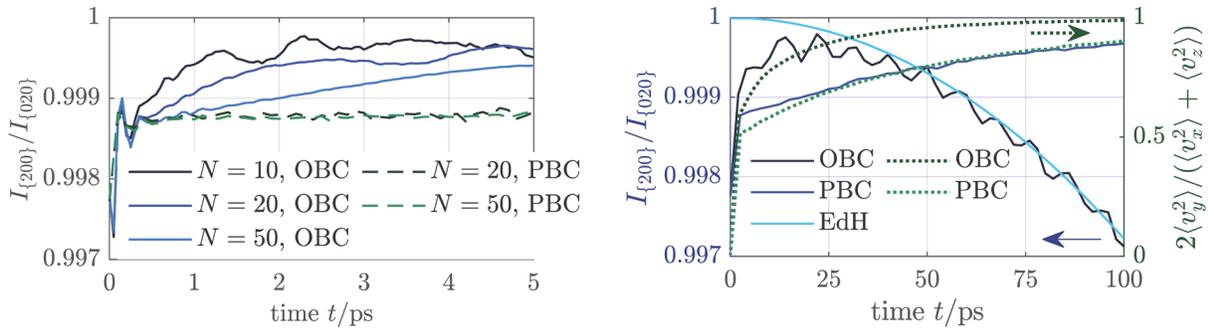

**Fig. S10:** Left panel: Finite-size effects of the anisotropy of crystallographically equivalent peaks comparing open boundary conditions (OBC) with periodic ones (PBC). For OBC a finite-size effect is observed: the relaxation time of the contrast increases system size. PBC do not show this effect. Right panel: Long-time evolution for $N = 50$ testing the three cases (OBC, PBC, and global rotation according to the Einstein-de Hass effect (EdH)) (blue, solid lines). Also shown is the anisotropy of the mean-squared velocities $2\langle v_y^2\rangle/(\langle v_x^2\rangle+\langle v_z^2\rangle)$ (green, dotted lines).

*Temperature dependence.* The initial conditions explained above are designed to model the angular momentum transfer from the spins to the lattice. The resulting lattice temperature in our simulation is $\Delta T = 15$ K, much lower as compared to the experiment[1]. To test the potential influence of an additional energy transfer from the electronic system to the lattice, we modify the initial conditions of our simulations such that we keep the angular momentum, but additionally impose a higher energy on the lattice by displacing all atoms that have not been excited with angular momentum by some amount in random directions. This can lead to a much higher temperature of, e. g., $\Delta T = 60$ K in Fig. S11. Nevertheless, we determine practically the same anisotropy contrast as compared to the previous case. We conclude that an initial temperature before laser excitation does not affect the anisotropy of our Bragg peaks and the role of the polarized phonons.

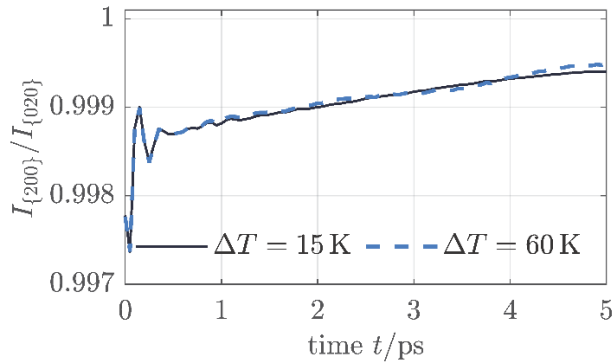

**Fig. S11.** Temperature dependence: The anisotropy of crystallographically equivalent peaks for the same angular momentum $\vec{L}_0$ (same demagnetization) but different energy transfers to the lattice, leading to a temperature increase of $\Delta T = 15$ K and 60 K, respectively.

## Magneto-optical Faraday effect

Figure S12 shows a measurement of the dynamical magnetization M in our single-crystalline thin-films via recording the magneto-optical Faraday effect as a function of laser excitation. The pump pulse parameters are made similar to the conditions of the diffraction experiment, including laser repetition rate, incidence angle, wavelength and pulse duration. Fluence is adjusted to below the damage threshold with a similar procedure as in the diffraction experiment. Figure S12a shows the measured magnetic hysteresis curves for a delay times before time zero (black) in comparison to a trace taken 7 ps after laser excitation (blue). We see the expected decrease of amplitude but no change of shape. Figure S12b shows an analysis of the time-resolved dynamics. The degree of demagnetization is 40-50 %, the dynamics is faster than 0.5 ps and demagnetization persists for tens of picoseconds. All results agree with earlier observations of similar materials at similar excitation fluences [26,27].

---

[1] Note, however, that the lattice temperature is not well-defined on short time scales, since equipartition is violated. The temperature presented here is defined via $\frac{3}{2} k_\text{B} T = \langle v_x^2 \rangle + \langle v_y^2 \rangle + \langle v_z^2 \rangle$ as a measure for the energy pumped into the system.

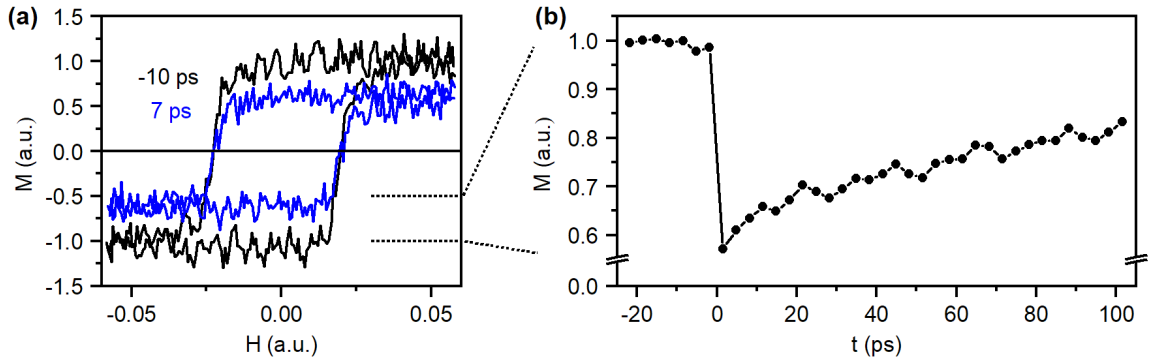

**Fig. S12.** Magneto-optical Faraday effect of our thin-film nickel specimen. (a) Measured hysteresis curves for a negative (black) and slightly positive pump-prober delay (blue). (b) Magnetization as a function of delay time.

**Bragg spot positions and crystal expansion**

On picosecond time scales, there is not enough time for our specimen crystal to change its lattice constants by volumetric effects. Figure S13 shows a measurement of the positions of the eight nickel spots of Fig. 1d as a function of the pump-probe delay time. All Bragg spot position changes stay below $2\times10^{-4}$. Our nickel crystal therefore remains in all of the investigated time range in its original volume without isotropic or anisotropic deformations.

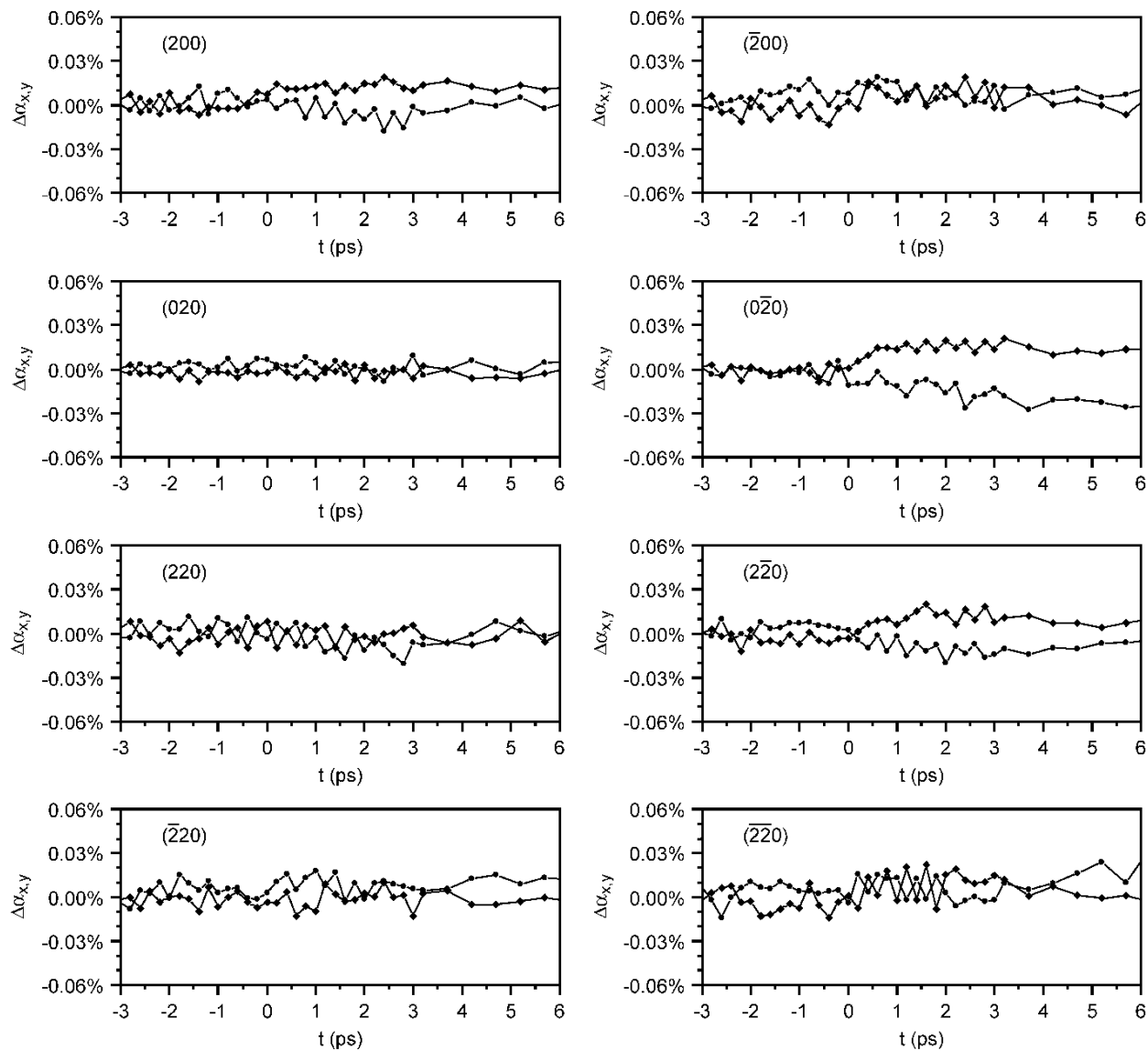

**Fig. S13.** Changes of Bragg spots angles as a function of pump-probe delay. Dots, changes $\Delta\alpha_x$ along the *x*-axis; squares, changes $\Delta\alpha_y$ along the *y*-axis (see Fig. 1d).

## Fluence dependence

Figure S14 shows a measurement of the ultrafast phonon dynamics as a function of the applied laser fluence. The left panel shows the Bragg spot intensity changes of {200} after integration of t > 2 ps (see Fig. 2a). The middle panel shows the measured Bragg spot anisotropy, analyzed in the same way as in Fig. 2b and then integrated for t > 2 ps. The last data points in both graphs are the results from Figs. 2a-b. The right panel shows the results of the molecular dynamics simulations as a function of the degree of demagnetization. Time traces are integrated between 1-2 ps for n = 50. As predicted by the analytical considerations in Eq. (1), the dependency is linear and the measured fluence data is consistent with this expectation.

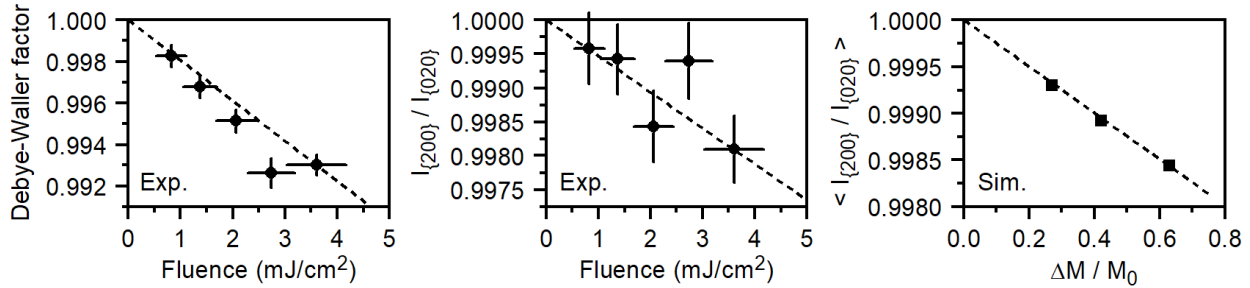

**Fig. S14.** Time-integrated Bragg spot responses as the Debye-Waller effect (left panel) and the anisotropy (central panel) as functions of the applied laser excitation fluence. The right panel shows the simulated anisotropy as function of the degree of demagnetization.

**Direct beam effects**

Demagnetization of the nickel membrane could potentially distort the trajectory of the electron beam and thereby produce unwanted diffraction effects. However, calculations of Lorentz forces predict a deflection of less than 5 μrad. In the experiment, we derive an upper limit by measuring electron beam deflections at high precision. We produce by two photoemission laser pulses a sequence of two electron pulses at a time difference of 30 ps. One of these electron pulses always passes through the specimen before the laser excitation and serves as a reference while the other electron pulse is the probe pulse of the experiment. The center positions of both beams are fitted with pseudo-Voigt functions. Recording the difference between these two centers cancels any potential drifts of the experiment and therefore produces a sensitive upper limit for magnetic beam deflections. Fig. S15a shows the two electron beams on the screen. Figure S15b reveals the Debye-Waller effect in the probe beam (blue) but not in the reference beam (black), as expected. Figure S15c shows the differences between of the two beam positions on the screen. All data remains below 5 μrad; this is more than $10^4$ times smaller than the rocking curve width. Dynamical electron beam effects are therefore not significant for the reported results.

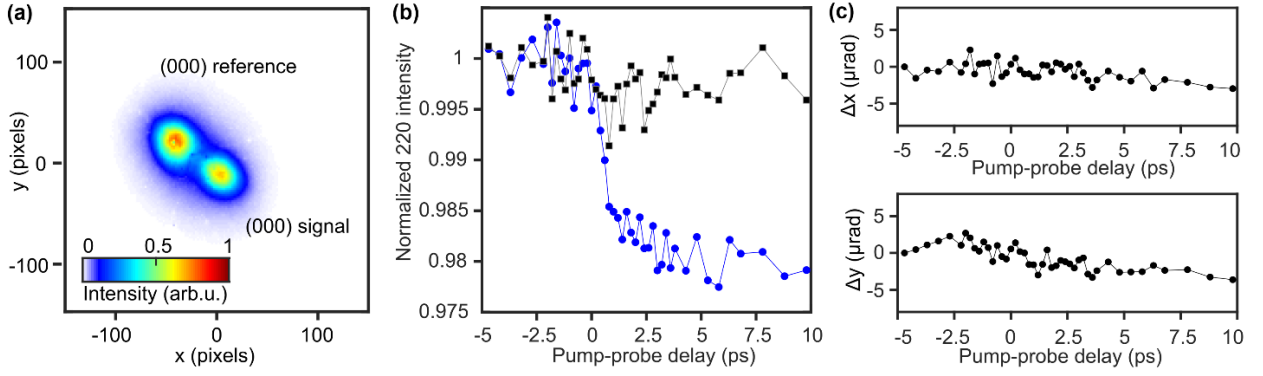

**Fig. S15.** (a) Two time-delayed electron beams on the screen. (b) Intensity changes in the reference pulse (black) and probe pulse (blue), showing a Debye-Waller effect in the probe beam only. (c) Differences of the beam positions before and after laser excitation as a function of the pump-probe delay, converted to angle changes at the specimen. All changes remain below 5 μrad.

**Potential anisotropic lattice potentials**

Here, we rule out that the observed Bragg spot anisotropy is caused by possible anisotropic changes of the lattice potential along the magnetic direction. Such a bond hardening or bond softening could influence the atomic fluctuations in an anisotropic way. To exclude the significance of such effects, we estimate all possible energy scales that could anisotropically alter the lattice potentials and then show that these are much smaller than the effects we have measured. The thermodynamic potential $G(T, H, S, \varepsilon)$ with temperature $T$, magnetic field $H$, magnetization direction $\mathbf{S} = (S_1, S_2, S_3)$ and strain tensor $\boldsymbol{\varepsilon}$ contains anisotropic magnetic contributions due to magnetostriction or other magnetic anisotropies [28]. For bulk nickel, the largest contribution to the magnetostriction is given by $G_{ms} = B_1(\varepsilon_{11}S_1^2 + \varepsilon_{22}S_2^2 + \varepsilon_{33}S_3^2) + \ldots$ with the magneto-elastic constant $B_1 \approx 1$ meV/atom [29]. Upon complete demagnetization, this anisotropic energy contribution would vanish. The strain $\varepsilon_{11}$ for saturated magnetization in (100) direction is given by the magnetostriction coefficient $\lambda \approx 4 \cdot 10^{-5}$ for crystalline nickel [28,30]. The resulting energy change is therefore $G_{ms} = B_1 \lambda \approx 4 \cdot 10^{-5}$ meV/atom. Even for a much larger strain of the order of 1% [31], the resulting energy contribution would still be smaller then $10^{-2}$ meV/atom. A second relevant energy scale comes from magneto-crystalline anisotropy. For nickel, the largest contribution is $K_1 \approx -10^{-2}$ meV/atom [32]. In thin films, there are contributions from surface or interface effects, reaching a level of up to $10^{-1}$ meV per interface atom for strained Ni films; see Table 3 in Ref. [32].

Both energy scales are by several orders of magnitude smaller than the energy required to explain the experimental results. The measured anisotropic contributions to the mean atomic displacement, inferred from the Bragg spots of Fig. 2, require an energy of at least ~4 meV/atom. This value is more than a factor of 100 larger than the maximum energy contributions that can be produced by magnetostriction or magneto-crystalline anisotropies. Such effects are therefore not significant for our results and conclusions.

**Additional references**